\documentstyle[11pt,epsf,psfig]{article}

\def\simlt{\mathrel{\hbox to 0pt{\lower 3.5pt\hbox{$\mathchar"218$}\hss}
      \raise 1.5pt\hbox{$\mathchar"13C$}}}
\def\simgt{\mathrel{\hbox to 0pt{\lower 3.5pt\hbox{$\mathchar"218$}\hss}
      \raise 1.5pt\hbox{$\mathchar"13E$}}}

\def\ref{\par\noindent\hangafter=1\hangindent=1.3cm}

\def\al2{AlII$\lambda$167.0}
\def\si2{SiII$\lambda$152.6}
\def\mg2{MgII$\lambda\lambda$279.6,280.3}

\def\Heii{He~{\sc ii}}
\def\Heiii{He~{\sc iii}}

\begin{document} 

\noindent{\LARGE \bf  A companion to a quasar at redshift 4.7 \\}

\vskip 0.5truecm

\noindent {\large \bf Patrick Petitjean$^{\star}$$^{\dag}$, 
                      Emmanuel P\'econtal$^{\ddag}$,
                      David Valls--Gabaud$^{\S}$    \& 
                      St\'ephane Charlot$^{\star}$} \\

\vskip 0.1truecm
\normalsize
\noindent $^{\star}$Institut d'Astrophysique de Paris -- CNRS, 
98 bis, Boulevard Arago, F-75014 Paris, France.\\[1mm]
\noindent $^{\dag}$DAEC, URA CNRS 173, Observatoire de Paris-Meudon,
F-92195 Meudon, France.\\[1mm]
\noindent $^{\ddag}$Centre de Recherche Astronomique de Lyon, CNRS UMR142,
9, Avenue Charles Andr\'e, \\
\indent F-69561 Saint-Genis-Laval, France. \\[1mm]
\noindent $^{\S}$URA CNRS 1280, Observatoire de Strasbourg, 
11, rue de l'Universit\'e, F-67000 Strasbourg, France.\\[1mm]

\normalsize

\vskip 4mm
\hrule
\vskip 4mm

\noindent {\bf 
There is a growing consensus that the emergence of quasars
at high redshifts is  related to
the onset of galaxy formation$^1$, suggesting that the detection of
concentrations of gas accompanying such quasars should provide
 clues about the early history of galaxies. Quasar companions have
been recently identified at redshifts up to $z \approx 3$ (refs 2-4).
Here we report observations of Lyman-$\alpha$ emission (a tracer of
ionised hydrogen) from the companion to a quasar at $z$=4.702, corresponding
to a time when the Universe was less than ten per cent of its present
age. We argue that most of the emission arises in a gaseous nebula 
that has been photoionised by the quasar, but an additional component
of continuum light --perhaps quasar light scattered from dust in the
companion body, or emission from young stars within the nebula-- appears
necessary to explain the observations. These observations may be indicative of
the first stages in the assembly of galaxy-sized structures.}

\vskip 0.5truecm

\par
The bright quasar BR~1202--0725 was discovered in the Automatic Plate 
Measuring optical survey for
quasars with $z$~$>$~4 (ref. 5, see also ref. 6). The emission redshift is 
uncertain because
Ly$\alpha$ absorption produced by a foreground metal line system
is present at a redshift $z$~=~4.687 (ref. 7) that is much higher 
(by $\sim$5000~km~s$^{-1}$) than the
redshift inferred from intermediate resolution spectroscopy$^{7,8}$ for
the C~{\sc iv} emission line of the quasar.

We used the integral field spectrograph TIGER$^{9}$
(Traitement Int\'egral des Galaxies par l'Etude de leurs Raies) mounted on 
the 3.60~m Canada-France-Hawai Telescope during a
windy run on 4--7 February 1995. The instrument samples the field of view 
through an
array of micro-lenses, each providing a spectrum. 
Because of the windy weather during the observations, only two exposures were
useful, corresponding to 2.3 hours of integration. The seeing 
was 0.7" FWHM (Full Width at Half Maximum) and the instrumental configuration
led to a spatial sampling of 0.61" in a circular field of view 
of 10" diameter centred on the quasar and a spectral sampling of 
3.4~\AA. The spectral resolution, measured as the FWHM of an arc line,
was 10~\AA, and the wavelength range was
5850--7100~\AA, corresponding to a Ly$\alpha$ redshift range of 3.81--4.84.
The data were reduced in a standard way.$^{10,11}$ 
The 1$\sigma$ noise in the spectra obtained through one lens
is 3$\times$10$^{-19}$~erg~cm$^{-2}$~\AA$^{-1}$ after sky subtraction.
We have then reconstructed 60~{\AA} wide narrow-band images. \par

The line of sight towards BR~1202--0725 contains a known damped
Ly$\alpha$ absorption system with an H~{\sc i} column density 
$N$(H~{\sc i})~$\sim$~4$\times$10$^{20}$~cm$^{-2}$ at $z$~=~4.383 (refs 7,12). 
Such systems are generally thought to arise in proto-galactic 
discs$^{13}$ and so indicate  potentially intensive star formation regions. However,
we do not detect any emission down to a 3$\sigma$ limit of
3. 10$^{-18}$~erg~cm$^{-2}$~\AA$^{-1}$arcsec$^{-2}$  at $z=4.383$  in a region of 
41$\times$41$h^{-1}_{75}$~kpc$^2$ (where $h_{75}$ is the Hubble
constant in units of 75~km~s$^{-1}$~Mpc$^{-1}$ and $q_{o}$~=~0.5)
centred on the quasar.
The limit reached, consistent with previous studies, 
confirms that damped Ly$\alpha$ systems are not strong 
Ly$\alpha$ emitters$^{14,15}$, either because of their small but significant
dust content$^{16}$ or their low duty--cycle of massive star 
formation$^{17}$. \par

In contrast, the image centred at 6940~\AA~ (see Fig.~1)
exhibits a significant (flux greater than 
3$\sigma$) North-West extension (region B) with the maximum  emission located 
2" away from the quasar (or 9$h^{-1}_{75}$~kpc) that
coincides spatially with an emission excess previously seen in broad-band
$VrI^{18}$$K^{19}$ and narrow-band filter images$^{20}$. 
The contribution of the quasar has been carefully subtracted,
and we have coadded the spectra of all lenses in which the extension was
detected (Fig.~2).
There is a strong line (7~$\sigma$) centred at 6932~\AA. The total flux,
integrated over the line, is
2.5$\pm$0.6$\times$10$^{-16}$~erg~s$^{-1}$~cm$^{-2}$.
Identifying the line with Ly$\alpha$ gives a redshift $z$~=~4.702.
The emission is spatially resolved and has a maximum north-east/south-west 
extension of 1.5" after correction from seeing effects.\par 

As we detect only one line, other potential identifications could be
H$\alpha$ at $z$~=~0.056, H$\beta$ at $z$~=~0.43
or [O{\sc ii}]$\lambda$3727 at $z$~=~0.86. The probability of gravitational
amplification is greatest in the latter two cases but
population synthesis models
indicate that the broad--band observations are inconsistent with
these redshifts. In particular, the $r-K$ colour characterises a 
steep spectrum
that is difficult to reconcile with the lower limit on the $B$ magnitude.
Strong absorption systems are present at $z$~=~1.7544 and 2.4421
(refs. 7,12)
in the quasar spectrum, but these redshifts are ruled out since no emission
line is expected at 6932~\AA. \par

The detected line coincides in wavelength with the peak of the
quasar emission. This is a strong indication that the line is, in fact,
Ly$\alpha$ emitted by a quasar companion.
Moreover, the blue wing of the line is sharp and coincides almost perfectly 
with the position of the first Ly$\alpha$ absorption line seen in
the quasar spectrum (see Fig.~3). This coincidence can be accounted
for if the same absorber extends from the quasar to region B, implying a 
projected size of at least 9$h^{-1}_{75}$~kpc.
\par

The Ly$\alpha$ flux is 1.7 times larger than the value obtained from
previous narrow-band filter imaging$^{20}$. A thorough check of the 
photometric quality of our observations using several standard stars
yields an uncertainty of less than 20\%. In addition, the flux
determined for the quasar is consistent with published data$^{8}$. 
 The line luminosity is 
2.1$\times$10$^{43}$$h^{-2}_{75}$~erg~s$^{-1}$ and contributes 70\% of the 
$r$-band emission. 
\par

The observed flux is similar to that observed for quasar companions 
at lower redshifts (refs 2,3,21) and much larger than the upper limits 
set by surveys for primeval galaxies in blank fields$^{22, 23}$. 
This suggests that the presence
of the quasar induces a peculiar interaction 
with the companion, either in the form of enhanced star formation
or because the ionising flux from the quasar
contributes to the excitation of the gas. In fact, the ionising
flux of the quasar, estimated from the observed $R$ magnitude and
assuming a standard spectral index $\alpha=-1.5$, is about
9.3$\times$10$^{56}$~$h^{-2}_{75}$ photons per second. Under case~B
recombination, the observed Ly$\alpha$ emission of the companion then
implies that the fraction of the quasar luminosity intercepted by the
gas is about 7$\times$10$^{-4}$. If we approximate the
cloud as a sphere of diameter 6~$h_{75}^{-1}$~kpc, corresponding
to a covering factor of 3$\times$10$^{-2}$, the
filling factor of the gas within the cloud is found to be over 0.02. The
true value could be even larger if some of the Ly$\alpha$ photons did not
escape from the nebula.\par

The interpretation of the extended emission near the quasar is constrained by
the wavelength and equivalent width of the emission line, its contribution
to the $r$ band flux, the upper limit on the $B$-band emission, and the four
$VrIK$ broad-band fluxes. These constraints taken simultaneously rule out
the possibility that the emission line is not Ly$\alpha$ at the redshift of
the quasar. We consider hereafter two extreme models:
a starburst galaxy and a gaseous nebulosity photoionised by the quasar. 
Examples of
both situations exist at lower redshifts ($z$~$\sim$~3; refs. 2,3,4).
Virtually any model galaxy in which the spectrum is dominated by young stars
is a reasonable match to the data. Fig.~4{\it a} shows a typical fit obtained
for a $1.1 \times10^7\,$yr old model galaxy with metallicity $0.1Z_\odot$ 
and a constant star formation rate of $13\,M_\odot\,{\rm yr}^{-1}$. In this
case, the onset of star formation would have occurred at $z_f\approx 4.79$ 
(again for $h_{75}$~=~1 and $q_0$~=~0.5). However, in starburst galaxy models
the fraction of the $r$-band flux contributed by Ly$\alpha$ emission
is always less than 50~\%, i.e., much smaller than observed. Therefore such
models alone cannot explain the data.

Alternative models of a pure photoionised nebula also present some 
shortcomings. If the $I$ and $K$ luminosities are assumed to be produced by
reprocessing of the quasar's ionising radiation alone (Fig.~4{\it b} shows
such a model in which the reflecting cloud has a metallicity of $0.1Z_\odot$),
the predicted Ly$\alpha$ emission is so strong that acceptable fits of the $r$
band flux can be obtained only after assuming that a substantial fraction of
Ly$\alpha$ photons are removed from the line of sight, because, for example, of
orientation effects or absorption by dust.$^{16}$ Arguing for a 
small filling factor of the cloud (see above)
would not affect this 
conclusion because this would imply similar changes in both the Ly$\alpha$ 
and continuum fluxes. The [O~{\sc ii}] $\lambda$3727 flux predicted for the
model in Figure~4{\it b}, 5$\times$10$^{-18}$~erg~cm$^{-2}$~s$^{-1}$,
is consistent
with previous limits.$^{6}$ The strong predicted N~{\sc v}
emission could have been marginally detected (assuming that the
cloud has zero metallicity would remove not only N~{\sc v} but also C~{\sc iv}
emission, hence worsening the fit of the $I$-band flux). More important, in
addition, is that the model cannot reproduce the observed $V$ magnitude.
Hence, although the Ly$\alpha$ emission is most probably a consequence of the
excitation by the ionising flux from the quasar, we conclude that an additional
source of continuum is likely to be present.\par
 
A possibility is that part of the continuum light from the quasar is scattered
in our direction, potentially by dust. In fact, a large amount of dust has
been detected in the quasar by continuum emission at millimetric and
sub-millimetric wavelengths.$^{24,25}$ New
observations suggest that the emission is elongated towards the companion (A.
Omont et al., mss in preparation), also indicating a large concentration of 
gas in that direction. We note that, by analogy with radio-galaxies, the 
ultraviolet continuum light of the companion could result from both 
scattered light and stellar emission.$^{26}$ The width of the Ly$\alpha$
line (more than 1000~km~s$^{-1}$, see Fig.~2) indicates strong disturbances
of the kinematics of the gas. This all together suggests that we are 
witnessing the first stages in the build-up of a galaxy-sized structure.
\par

\vskip 4.0cm
\small

We thank S. D'Odorico, S. Djorgovski and A. Omont for providing us
with information prior to publication. We also thank P.~Madau for
computing the transmission curve at $z=4.7$. The observations have
been collected at the Canada-France-Hawaii Telescope, which is operated
by CNRS of France, NRC of Canada, and the University of Hawaii.

\newpage
\vskip 5mm
\small

\noindent {\bf References} \\

\par\noindent
1. Haehnelt, M. \& Rees M.J. {\sl MNRAS} {\bf 263}, 168--178 (1993)
\par\noindent
2. Djorgovski, S., Spinrad, H., McCarthy P. \& Strauss, M.A. 
{\sl Astrophys. J.} {\bf 299}, L1--L5 (1985)
\par\noindent
3. Steidel, C.C., Sargent, W.L.W. \& Dickinson, M. {\sl Astron. J.} 
{\bf 101}, 1187--1195 (1991)
\par\noindent
4. M{\o}ller, P. \& Warren, S.J. {\sl Astr. Astrophys.} {\bf 270}, 43--52 
(1993)
\par\noindent
5. Irwin, M.J., McMahon, R.G. \& Hazard, C. in {\sl The Space Distribution of
Quasars} (ed Crampton, D.) {\bf 21}, 117--126 (ASP Conf. Ser., 1991)
\par\noindent
6. Pahre, M.A. \& Djorgovski, S.G. {\sl Astrophys. J.} {\bf 449}, L1--L4
(1995)
\par\noindent
7. Wampler, E.J., Williger, G.M., Baldwin, J.A., Carswell, R.F., Hazard,
C. \& McMahon, R.G. {\sl Astr. Astrophys.}, in the press (1996)
\par\noindent
8. Giallongo, E., D'Odorico, S., Fontana, A., McMahon, R.G., Savaglio, S.,
Cristiani, S., Molaro, P. \& Trevese, D. {\sl Astrophys. J.} {\bf 425}, L1--L4
(1994)
\par\noindent
9. Bacon, R. et al. {\sl Astr. Astrophys. Suppl. Ser.} {\bf 113}, 347-357
(1995) 
\par\noindent
10. Ferruit, P. \& P\'econtal, E. {\sl Astron. Astrophys.} {\bf 288}, 65-76 
(1994)
\par\noindent
11. Bacon, R., Emsellem, E., Monnet, G. \& Nieto, J.L. {\sl Astr. Astrophys.} 
{\bf 281}, 691--712 (1994)
\par\noindent
12. Lu, L., Sargent, W.L.W., Womble, D.S. \& Barlow, T.A. {\sl Astrophys. J.} {\bf 457}, L1 (1996) 
\par\noindent
13. Wolfe, A.M. {\sl Astrophys. J.} {\bf 402}, 411--419 (1993)
\par\noindent
14. Pettini, M., Hunstead, R.W., King, D.L. \& Smith, L.J. in {\sl QSO
Absorption Lines} (ed Meylan, G.) 55--58 (ESO, 1995)
\par\noindent
15. Lowenthal, J.D. et al. {\sl Astrophys. J.} {\bf 451} 484--497 (1995)
\par\noindent
16. Charlot, S., \& Fall, S.M. {\sl Astrophys. J.} {\bf 415}, 580--588
(1993)
\par\noindent
17. Valls--Gabaud, D. {\sl Astrophys. J.} {\bf 419}, 7--11 (1993)
\par\noindent
18. Fontana, A., Cristiani, S., D'Odorico, S., Giallongo, E. \& Savaglio, S.
{\sl Mon. Not. R. astr. Soc.}, in the press (1996) 
\par\noindent
19. Djorgovski, S.G. in {\sl Science with VLT} (eds Walsh, J.R.  \& Danziger,
J.) 351--360 (ESO, 1995)
\par\noindent
20. Hu, E.M., McMahon R.G. \& Egami E. {\sl Astrophys. J. Letters}, in the 
press (1996)
\par\noindent
21. Hu, E.M., Songaila, A., Cowie, L.L. \& Stockton, A. {\sl Astrophys. J.}
{\bf 368}, 28--39 (1991) 
\par\noindent
22. Thompson, D., Djorgovski, S. \& Trauger, J. {\sl Astr. J.} 
{\bf 110}, 963--981 (1995)
\par\noindent
23. Pritchet, C.J. {\sl PASP} {\bf 106}, 1052--1067 (1994)
\par\noindent
24. McMahon, R.G., Omont, A., Bergeron, J., Kreysa, E. \& Haslam, C.G.T. 
{\sl Mon. Not. R. astr. Soc.} {\bf 267}, L9--L12 (1994)
\par\noindent
25. Isaak, K.G., McMahon, R.G., Hills, R.E. \& Withington, S. {\sl
Mon. Not. R. astr. Soc.} {\bf 269}, L28--L32 (1994)
\par\noindent
26. McCarthy, P.J., {\sl Ann. Rev. Astron. Astrophys.} {\bf 31}, 639-688
(1993)
\par\noindent
27. Ferland, G.J. University of Kentucky Department of Physics \& Astronomy 
Internal Report (1993)
\par\noindent
28. Madau, P. {\sl Astrophys. J.} {\bf 441}, 18--27
(1995)
%\par\noindent
%\bigskip
%\par\noindent

%-------------- Figures

\newpage
\pagestyle{empty}
\begin{figure*}[h]
 \protect \centerline{
\psfig{file=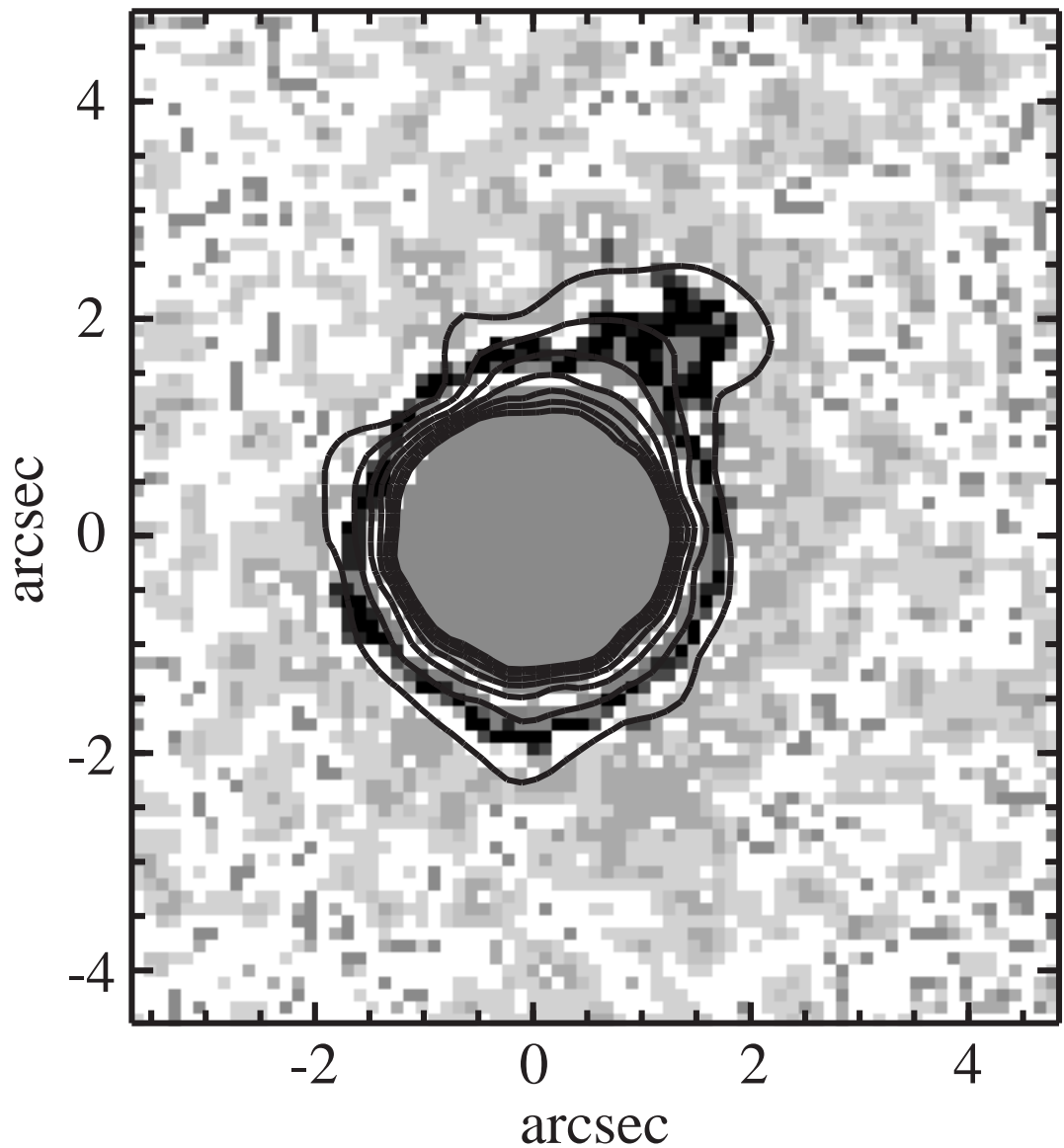,width=23.0cm,angle=90.}}
\caption{  Isophotes of the emission detected in the wavelength
range 6910--6970~\AA~ plotted over the $r$-band image$^{18}$.
A Gaussian spatial 
filter was applied to the merged data cube, leading to a final spatial 
resolution of 1.2" (measured as the FWHM of the quasar image in the
reconstructed images). 
The emission detected NW of the quasar peaks at 2" from the quasar.
Contours of Ly$\alpha$ surface brightness are
0.5, 1.2, 1.9, 2.6, 3.3, 4.0 and 4.7 
10$^{-16}$~erg~s$^{-1}$~cm$^{-2}$~arcsec$^{-2}$. } 
\label{f1}
\end{figure*}

\newpage
\pagestyle{empty}
\begin{figure}[t]
\protect \centerline{
 \psfig{file=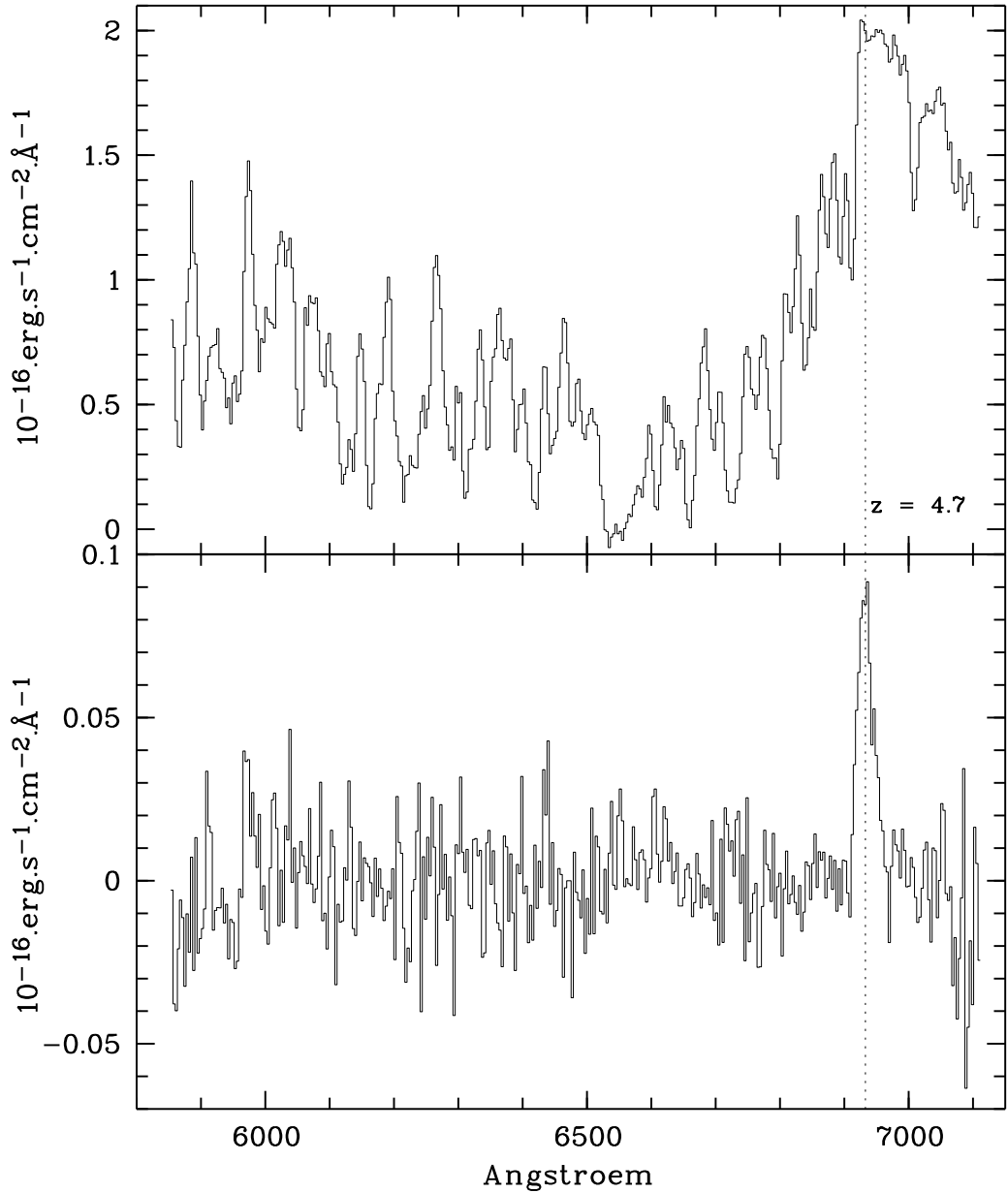,width=15.0cm}}
\caption{ Spectra of the quasar (upper panel) and region B (lower panel).
The position of a $z$~=4.7 Ly$\alpha$ line is indicated by a vertical dashed
line. The quasar spectrum 
has been obtained by adding the spectra in all 
lenses in which its intensity is larger than three times the rms noise
in the reconstructed broad band image. The quasar spectrum has been 
subtracted 
from the spectra of the lenses sampling region B.
As the companion is detected only in the Ly$\alpha$ line, the
continuum map is an image of the unresolved quasar and can thus be used  as 
the weight 
map  to be applied  to remove the quasar contribution everywhere in the field.
The spectra of all lenses have then been coadded. } 
\label{f2}
\end{figure}

\newpage
\pagestyle{empty}
\begin{figure*}[t]
\protect \centerline{
 \psfig{file=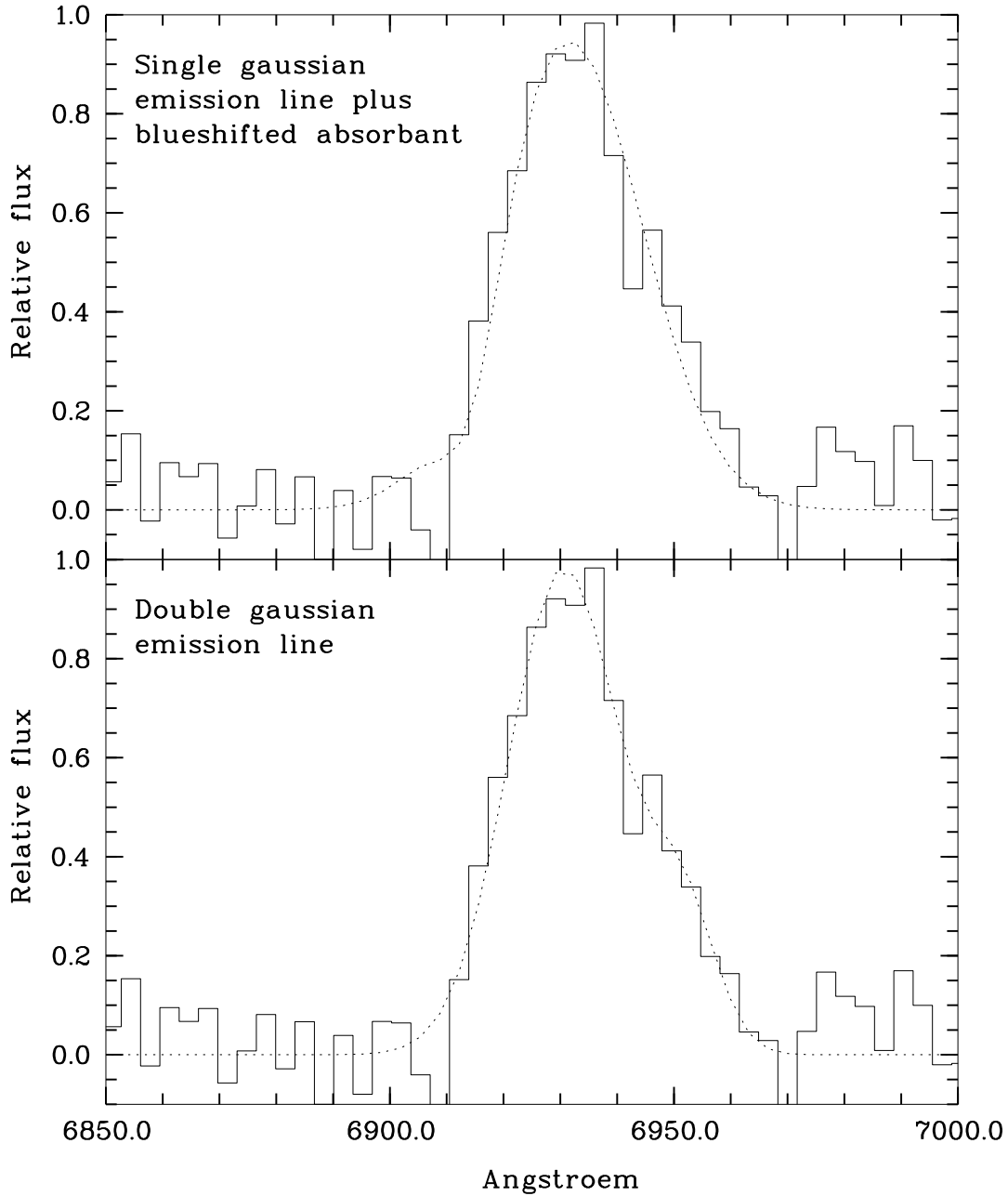,width=15.0cm}}
\caption{ Model fits of the emission line detected in region B (dashed
lines) overplotted on the spectrum (solid line). 
({\it a}) The model includes a single Gaussian profile absorbed in the 
blue by an intervening absorber. The emission profile peaks at
$\lambda$~=~6931.5~\AA~ and has FWHM~=~1245~km~s$^{-1}$. The absorption
line is centred at 6914.5~\AA~ and has FWHM~=~6~\AA. The reduced $\chi^2$ 
is 1.04. ({\it b}) The model includes two Gaussian profiles 
centred at 6930.7 and 6951.4~\AA~ (separation 890~km~s$^{-1}$)
and with FWHM~=~900~km~s$^{-1}$ and 410~km~s$^{-1}$ respectively. The reduced 
$\chi^2$ is 0.96. } 
\label{f3}
\end{figure*}

\newpage
\pagestyle{empty}
\begin{figure*}[h]
\hfil \epsfxsize 16cm\epsfbox{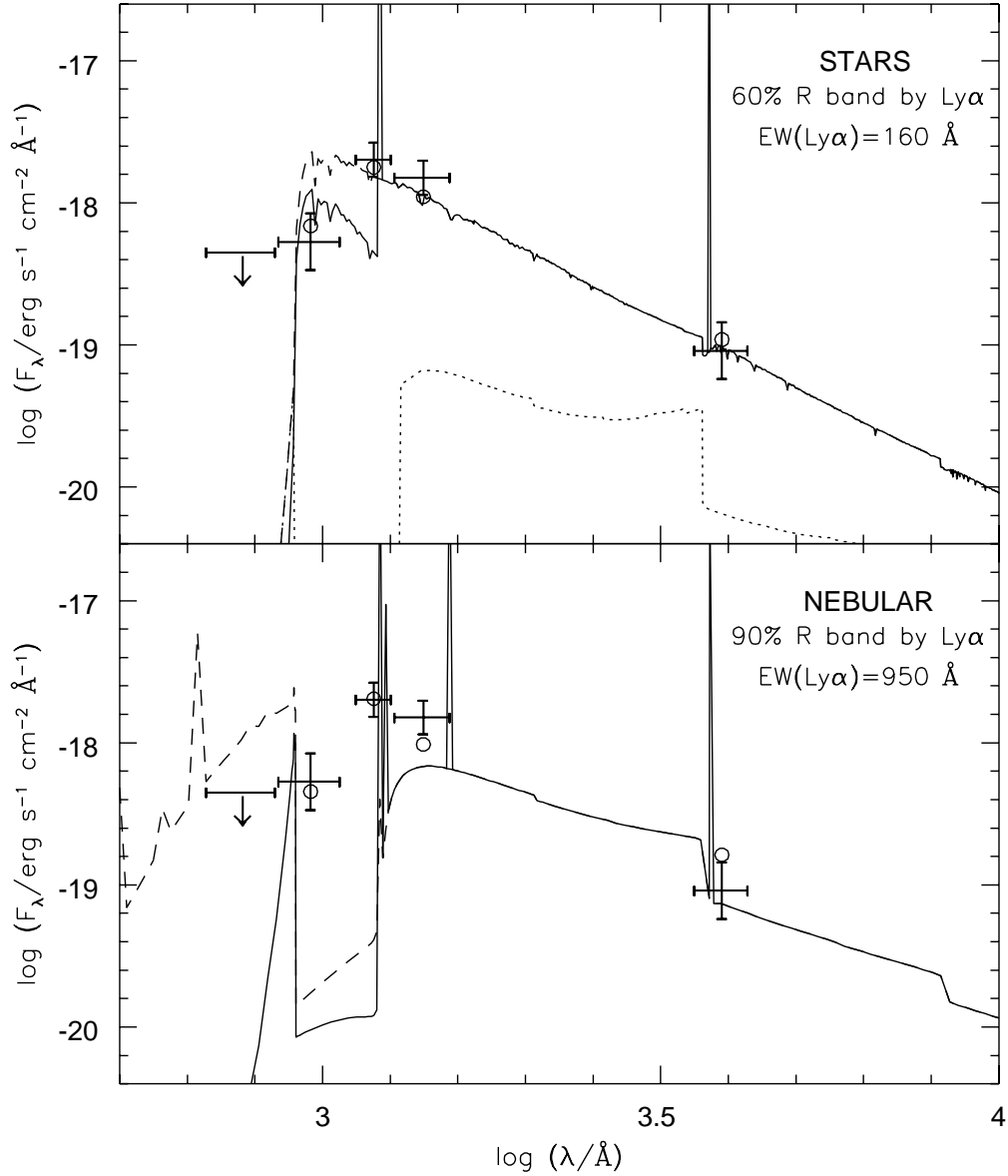}\hfil
\caption{ Model fits to the rest-frame emission properties
of region B. Vertical error bars represent uncertainties in the broad-band 
observations, and horizontal error bars the filter bandwidths. Open circles
represent filter-averaged model fluxes. Also indicated in each panel is the
contribution of the Ly$\alpha$ line to the integrated $r$-band flux.
({\it a}) Spectrum of a $1.1 \times10^7\,$yr old galaxy with metallicity
$0.1 Z_\odot$, a Salpeter initial mass function with lower and upper cutoffs
0.1 to 100$\,M_\odot$, respectively, and a constant star formation rate
of $13\,M_\odot\, {\rm yr}^{ -1}$, as predicted by recent population synthesis
models (G. Bruzual \& S.C., mss in preparation). 
Nebular emission was computed$^{27}$ assuming case~B recombination
and gas parameters typical of giant H{\sc ii} regions 
($n_e=100\, $cm$^{-3}$, $T_e
=10^4 \,$K, $n$(\Heii)/$n$(H)=0.1, and $n$(\Heiii)/$n$(H) 
=0.01). The dotted line shows the resulting nebular continuum and the
solid line the sum of the stellar and nebular continua with superimposed 
Ly$\alpha$ and [O~{\sc ii}] emission lines. Blanketing of the radiation 
blueward of
Ly$\alpha$ and photoelectric absorption of the (nebular) radiation blueward
of the Lyman limit by foreground H{\sc i} absorbers at redshifts $z<4.687$ 
have been
included using approximate analytic formulae.$^{28}$ The dashed line shows
the total (stars plus nebular) spectrum in the absence of H{\sc i} absorption. 
({\it b}) Spectrum reflected by a cloud with metallicity $0.1Z_\odot$ 
photoionised by the quasar according to standard photoionisation 
models.$^{27}$ 
The ionising spectrum of the quasar has been approximated by a power 
law $f_\nu\propto\nu^{-1.5}$. The electron density, Lyman limit optical
depth, and ionisation parameter of the model are $n_e=100\, $cm$^{-3}$, 
$\tau_L=100$, and log $U=-1$, respectively, but the quality of the fit 
depends
only weakly on these parameters. To improve the fit, Ly$\alpha$ emission has
been reduced arbitrarily by 90\%. As in ({\it a}), the dashed and solid lines
show the spectrum before and after inclusion of photoelectric
absorption and Ly$\alpha$ blanketing by foreground H{\sc i} clouds. } 
\label{f4}
\end{figure*}

\end{document}